\documentstyle[12pt]{article}

\def \hL {\widehat L}
\def \F {F}
\def \my {|y|}
\def \nl {{\bar \eta}_{+}^1}
\def \nr {\eta_{+}^1}
\def \nn {((\nl \Gamma^p d\nr)\Gamma_p)}
\def \p {\partial}
\def \dx {(dx^p\Gamma_p)}
\def \dy {(dy^q\Gamma_q)}

\begin{document}
\begin{flushright}
hep-th/0002030\\
ADP-00-09/M89
\end{flushright}

\vspace{6mm}
\begin{center}
{\Large {\bf D1 and D5-Brane Actions in $AdS_m\times S^n$}}
\vspace{6mm}

{\Large P. Dawson}
\vspace{6mm}

Department of Physics and Mathematical Physics, University of
Adelaide, SA 5005, Australia.
 
email: pdawson@physics.adelaide.edu.au 
\vspace{6mm}

\rule{150mm}{.5mm}

\end{center}

\begin{abstract}
The $\kappa$-invariant and supersymmetric actions of D1 and D5-branes
in $AdS_3\times S^{3}$ are investigated, as well as
the action of a D5-brane in an $AdS_5\times S^{5}$ background.  The
action of a D5-brane lying totally in an $AdS_3\times S^{3}$
background is found.  Some progress was made towards finding the 
action for the D5-brane free to
move in the whole $AdS_3\times S^{3}\times T^{4}$ space, however 
the supersymmetric action found here is not
$\kappa$-invariant and the reasons the method used did not find a 
$\kappa$-invariant solution are discussed.  
\end{abstract}

\section{Introduction}

Since late 1997 there has been a lot of interest in superstring theory
in $AdS_m\times S^n$ space, as according to the Maldacena conjecture
\cite{jm,ew} string theory on $AdS_m\times S^n$ can be compared to
Conformal Field Theory in $m-1$ dimensions.  Initially, the greatest
interest fell to {\it IIB} superstring theory on $AdS_5\times S^5$, due to
$AdS_5\times S^5$ being a maximally supersymmetric vacuum of {\it IIB}
supergravity and its CFT counterpart being ${\cal N}=4$, $D=4$ Super Yang Mills
Theory.  However the problems of quantising this theory so that it can
be compared to it's CFT counterpart and of our poor
understanding of 4 dimensional CFT has led to much effort going into
the study of string theory on $AdS_3\times S^3\times {\cal M}^4$, which
should be simpler to compare to CFT. 

The $AdS_3\times S^3\times {\cal M}^4$ background is the near horizon
geometry of a {\it IIB} D1 - D5-brane system.  Finding the action of a D1 or
D5-brane in $AdS_3\times S^3\times {\cal M}^4$ is therefore like finding the 
action of one of the parallel and coincident D1 or D5-branes which cause 
the space to deform into $AdS_3\times S^3\times {\cal M}^4$, if the relevant D1
or D5-brane has been slightly extracted from the rest of the D-branes.
Here ${\cal M}^4$ is a Ricci-flat four dimensional compact
manifold, such as $K3$ or $T^4$ \cite{jm,ew}. 

Despite this being a much studied system there has been a gap in the
research so far, ie: finding the full $\kappa$-invariant and
supersymmetric {\it IIB} effective actions of the D1 and D5-branes in
$AdS_3\times S^3$ (although some papers have either: proposed a method
for finding the D1-brane action \cite{pst}; given just
the BI-action \cite{ckktvp}; given a bosonic action and performed
an energy analysis of the branes \cite{r,grst,4p}; or used the BPS method to
find a D5-brane BPS solution of supergravity in $AdS_5\times
S^5$\cite{cgmp}).  

In this paper we determine the action of the D1 and D5-branes in
an $AdS_3\times S^3$ background, where both the D1 and D5-branes lie
entirely in $AdS_3\times S^3$ and the ${\cal M}^4$ space has been
compactified to zero volume.  This D5-brane does not correspond to one of
the D5-branes that are deforming the background to $AdS_3\times
S^3\times {\cal M}^4$, as such branes have 4 dimensions in ${\cal
M}^4$.  The method used was based on the method of Metsaev and Tseytlin in
\cite{mt}.  The actions are described by constructions of Cartan forms
on the coset superspace.  The supersymmetric BI-action is found first,
and then $\kappa$-symmetry and supersymmetry are used to determine
what WZ term is needed to complete the action
\cite{cgnw,cgnsw,c,aps,bt,bst}.  The coset superspace of $AdS_3\times S^3$ is
$\frac{SU(1,1|2)^{2}}{SO(2,1)\times SO(3)}$ which describes the
spacetime supersymmetries the action should display.  

The D5-brane found above is of interest due to the fact that as the
brane must lie in $AdS_3\times S^3$, instead of the larger space
$AdS_3\times S^3\times {\cal M}^4$, this can be
interpreted as a constraint imposed on the fields of a D5 action in
$AdS_3\times S^3\times {\cal M}^4$, such that the solution must lie in
$AdS_3\times S^3$ and with the ${\cal M}^4$ compactified away.  This
constraint on the solutions can be thought
of as
describing a brane that is lying in the background of
another D5-brane (or branes), with the brane dimensions in ${\cal
M}^4$ compactified away and with the branes intersecting in one
spatial dimension.  Such a configuration is S-dual to the NS5-NS5
configuration described in \cite{at}, which is BPS and thus such a
D5-D5 system is also BPS.

Next the action of the D5-brane in $AdS_5\times S^5$ was found using
the background coset superspace of
$AdS_5\times S^5$, $\frac{SU(2,2|4)}{SO(4,1)\times SO(5)}$.  This
was initially done as an exercise in preparation for handling a D5-brane in
$AdS_3\times S^3\times {\cal M}^4$, however it is of interest in its own
right.  This brane,
described by the BPS method in \cite{cgmp}, is related to the much
studied D5-brane of \cite{ew2,grst,yi,4p} (and others), which describes a
D5-brane in the presence 
of N D3-branes, and connected to them by N fundamental strings.  Such
a brane describes a baryon vertex in the corresponding 4 dimensional
gauge theory.  To transform the D5-brane action found here to the
D5 action of \cite{cgmp,ew2,grst,yi,4p} restrictions \cite{grst,yi}
must be placed on 
which dimensions the D5-brane can lie in and some of the fermionic
degree of freedom must be projected out to accomodate the presence of
the fundamental strings as well as the D5-brane.

Finally an attempt was made to find the most general D5-brane action in
$AdS_3\times S^3\times T^4$.  The action found using the method here is not
fully $\kappa$-invariant and the reason why this is so is discussed.


This paper has been structured as follows.  In subsection 2.1 Cartan
forms are introduced, and it is described how these objects can be
used to describe D-brane actions.  The $\kappa$-invariance constraint
that ensures that the D-brane does not have too many fermionic degrees
of freedom is also
discussed.  In subsection 2.2 the supersymmetric field strength $F$ is
defined.  In Sections 3 and 4 the actions of the D1-brane and D5-brane
in $AdS_3\times S^3$
are found, respectively.  In Section 5 the slightly more complicated
case of a D5-brane action in $AdS_5\times S^5$ is found.  Section 6
describes the attempt to find a fully supersymmetric and
$\kappa$-invariant action in $AdS_3\times S^3\times T^4$.  Section 7 is a
summary of the results.  Appendix A contains all the algebra and
notation used in the three backgrounds, and Appendix B contains the
Maurer-Cartan equations and other useful Cartan form
relations. Appendix C contains the gauge fixed actions of the D1 and
D5-branes, using the Killing gauge of \cite{kr}.


\section{Cartan 1 Forms and the Supercoset Method}

\subsection{Cartan Forms and the Action}

The notations and algebra used when finding actions on the three
different backgrounds $AdS_3\times S^3$, $AdS_5\times S^5$ and
$AdS_3\times S^3\times T^4$ can be found in Appendix A.  They have
been formulated in such a way that the resulting Maurer-Cartan
Equations, the Cartan form variations and various other useful
relations can be written out in the same way for the three different
backgrounds.  The notation used is based upon that in \cite{mt,o}.

The papers \cite{cgnw,cgnsw,ds} contain useful reviews of
the embedding of brane coordinates in background coordinates for use
in deriving the brane actions.

As in \cite{mt,o}, for a particular coset superspace the left invariant
Cartan 1 forms are\\
\begin{center}
$L^{A}=dX^{M}L_{M}^{A}$, \ \ \ \ $X^{M}=(x,\theta)$ are the background
super coordinates and\\

\begin{equation}
G^{-1}dG=L^{\hat{a}}P_{\hat{a}}+
\textstyle{\frac{1}{2}}L^{\hat{a}\hat{b}}J_{\hat{a}\hat{b}}+L^{\hat{\alpha}}
Q_{\hat{\alpha}}
\end{equation}
\end{center}
where $G=G(x,\theta)$ is a coset representative of
$\frac{SU(1,1|2)^2}{SO(2,1)\times SO(3)}$ in
$AdS_3\times S^3$, $\frac{SU(2,2|4)}{SO(4,1)\times SO(5)}$ in $AdS_5\times S^5$ and
$\frac{SU(1,1|2)^2}{SO(2,1)\times SO(3)}\times U(1)^4$ in
$AdS_3\times S^3\times T^4$.

The Cartan forms $L^{A}$ are inherently supersymmetric.  They are also
defined such that there is no mixing between the $AdS_m$, $S^n$ and ${\cal
M}^4$ dimensions of space (ie: $L^{a a'}=L^{a a''}=L^{a' a''}=0$).  In
the rest of this paper, the spinor index $\hat{\alpha}$ in
$L^{\hat{\alpha}}$ is not explicitly written.

As in \cite{mt}, the gauge choice $G(x,\theta)=g(x)e^{\theta Q}$ is
chosen, needed to explicitly find the supersymmetric field strength
$F$ and the R-R fields in $S_{WZ}$.  $g(x)$ is a coset representative
of $\frac{SO(2,2)\times SO(4)}{SO(2,1)\times SO(3)}\cong
\frac{SO(2,1)^2\times SO(3)^2}{SO(2,1)\times SO(3)}$ \cite{rr} in $AdS_3\times
S^3$, $\frac{SO(4,2)\times SO(6)}{SO(4,1)\times SO(5)}$ in $AdS_5\times
S^5$ and $\frac{SO(2,1)^2\times SO(3)^2}{SO(2,1)\times SO(3)} \times
U(1)^4$ in $AdS_3\times S^3\times T^4$.  $g(x)$ satisfies

\begin{equation}
g(x)^{-1}dg(x)=e^{\hat a}P_{\hat a} + \frac{1}{2}\omega^{{\hat a}{\hat
b}}J_{{\hat a}{\hat b}}.
\end{equation}

The method used is to find the supersymmetric BI-action and to then
use $\kappa$-symmetry to find the WZ-action.  The total effective action
of a Dp-brane can then be written as
\begin{eqnarray}
& &S=S_{BI}+S_{WZ}\\
& &S_{BI}=-\int_{M_{p+1}} d^{p+1}\sigma\ \sqrt{-det(G_{ij}+\F_{ij})}\\
& &G_{i j}=L_{i}^{\hat{a}} L_{j}^{\hat{b}}
\eta_{\hat{a}\hat{b}}=\partial_{i}X^{M}L_{M}^{\hat{a}}
\partial_{j}X^{N}L_{N}^{\hat{b}}
\eta_{\hat{a}\hat{b}}\hspace{2cm} L^{\hat a}=d\sigma^i L_{i}^{\hat a}
\\
& &S_{WZ}=\int_{M_{p+1}}e^{-\F}\wedge C, \hspace{2cm}
C=\bigoplus_{n=0}^{p+1}C_{n} 
\end{eqnarray}
where $C_n$ are the R-R super-$n$-forms and the brane tension has been set to
1 \cite{cgnw,cgnsw,c,aps,bt,bst}.

The general form of $d(e^{-F}\wedge C)$ can be deduced from \cite{cgnsw}
which endeavours to find the D-brane actions on a general background.
However extra terms are needed for the WZ-actions here to make them
fully $\kappa$-symmetric.

The $\kappa$-transformation
\begin{equation}
\delta_{\kappa}x^{\hat a}=0,\ \ \ \ \delta_{\kappa}x^{{\hat a} {\hat
b}}=0,\ \ \ \ \delta_{\kappa}\theta=\kappa,\ \ \ \
\frac{1}{2}(1+\Gamma)\kappa=\kappa, 
\end{equation}
is applied to the BI-action and then the WZ-action is defined such
that it cancels the $\kappa$ variation of the BI-action.  $\Gamma$
must have $\pm 1$ eigenvalues and $\Gamma^2=1$.  Thus
$\frac{1}{2}(1+\Gamma)$ projects out half of the fermionic degrees of
freedom, as needed for supersymmetry. Also, $d^{p+1}\sigma\ \Gamma$ must be a
$p+1$ form in order to relate $\delta_{\kappa}S_{BI}$ and
$\delta_{\kappa}S_{WZ}$. 

The $\Gamma$ for a D$p$-brane is \cite{cgnsw,mt,cgnw,c,bt,bst}
\begin{equation}
d^{p+1}\sigma\ \Gamma=(-1)^{n}\frac{e^{-\F}}{-\cal{L}_{BI}}\wedge
\bigoplus_{n}\Gamma_{(2n)}{\cal K}^{n}{\cal E} |_{vol}, \hspace{1cm}
\Gamma_{(2n)}=L^{\hat{a}_{1}}\Gamma_{\hat{a}_{1}}\cdots
L^{\hat{a}_{2n}}\Gamma_{\hat{a}_{2n}}, \label{dsigma}
\end{equation}
where ${\cal K}$ and ${\cal E}$ are defined in (\ref{Q}).

\subsection{Defining $F$}
As in \cite{mt}, $F$ is defined using the gauge choice
$G(x,\theta)=g(x)e^{\theta Q}$.
\begin{equation}
\F=dA+2i\int_{0}^{1}dt\ \bar{\theta}\hL_{t}\wedge {\cal K}
L_{t},\hspace{2cm} \F=\frac{1}{2}d\sigma^i\wedge
d\sigma^j\F_{ij}, \hspace{2cm} \hL=L^{\hat a}\Gamma_{\hat a},
\label{F}
\end{equation}
where $L_{t}^{\hat a}(x,\theta)=L^{\hat a}(x,t\theta)$,
$L_{t}(x,\theta)=L(x,t\theta)$.  This definition holds in all the three
backgrounds used in this paper, due to the choice of the notation used
(Appendix A).

It is not obvious that $F$ is
supersymmetric, as it is not defined entirely by Cartan forms.  It,
however, can be shown that
\begin{equation}
d\F=i\bar{L}\wedge \hL\wedge{\cal K} L,
\end{equation}
which is supersymmetric.  This is shown using equations
(\ref{dL1}), (\ref{dL2}), (\ref{pL1}), (\ref{pL2}) and the
Fierz identity \cite{o}
\begin{equation}
(\bar{A}\Gamma_{\hat{a}}B)(\bar{C}\Gamma^{\hat{a}}D)=
-\frac{1}{2}(\bar{A}\Gamma_{\hat{a}}e_{l}D)(\bar{B}\Gamma^{\hat{a}}e_{l}C)+
\frac{1}{2}(\bar{A}\Gamma_{\hat{a}}e_{l}C)(\bar{B}\Gamma^{\hat{a}}e_{l}D),
\label{Fierz1} 
\end{equation}
where $A,B,C,D$ are fermionic 0-forms, and $e_{l}=\{{\bf 1},{\cal E,J,K}\}$.

Thus, $F$ is supersymmetric if the variation of $A$ is defined such that it
cancels the variation of the second term in $\delta F$ \cite{cgnw,cgnsw,aps}.

\setcounter{equation}{0}

\section{D1-brane in $AdS_3\times S^3$}

The D1-brane BI-action is
\begin{equation}
S_{BI}=-\int_{M_{2}} d^2\sigma\ \sqrt{-det(G_{ij}+\F_{ij})},
\end{equation}
where
$i,j\;\;\in\;\{0,1\}$ are brane coordinates.  This is supersymmetric
but not $\kappa$-invariant. 

From (\ref{dsigma}) the operator $\Gamma$ which defines the $\kappa$
variation for this brane is
\begin{equation}
\Gamma=\frac{\varepsilon^{i_1 i_2}(\Gamma_{i_1 i_2}{\cal J} + \F_{i_1
i_2}{\cal E})}{2 {\cal L}_{BI}}.
\end{equation}
Using 
\begin{equation}
\delta det(G+\F)=det(G+\F)Tr((G+\F)^{-1}(\delta G+\delta\F)),
\end{equation}
one can show that
\begin{equation}
\delta_{\kappa}S_{BI}=
2i\int_{M_2}\bar{\kappa}\hL\wedge{\cal
J}L.
\end{equation}
Next, one shows that the $\kappa$-variation of the following supersymmetric
WZ-action

\begin{equation}
S_{WZ}=-2i\int_{M_2}\int_{0}^{1}dt\ \bar{\theta}\hL_{t}\wedge{\cal
J}L_{t}\ =-i\int_{M_3}\bar{L}\wedge\hL\wedge{\cal
J}L \label{WZ}
\end{equation}
cancels this variation.  Equivalence of the two different forms of
$S_{WZ}$ can be seen via

\begin{eqnarray}
S_{WZ} & = & \int_{M_3}\int_{0}^{1}dt\
\partial_t(-i\bar{L}_{t}\wedge\hL_{t}\wedge{\cal
J}L_{t}) \nonumber \\ 
& = & \int_{M_3}\int_{0}^{1}dt\
d(-2i\bar{\kappa}\wedge\hL\wedge{\cal
J}L)\\
& = & \int_{M_2}\int_{0}^{1}dt\ -2i\bar{\theta}\hL_{t}\wedge{\cal
J}L_{t}, \nonumber
\end{eqnarray}
where (\ref{dL1}), (\ref{dL2}), (\ref{pL1}), (\ref{pL2}), (\ref{initial}) and
(\ref{Fierz1}) were used. 

$S_{WZ}$ is invariant under supersymmetry trasformations as the second
form of the equation in (\ref{WZ}) is composed entirely of Cartan forms.

The full supersymmetric, $\kappa$-invariant D1-brane action in
$AdS_3\times S^3$ is thus given by
\begin{equation}
S^{D1}= -\int_{M_{2}} d^2\sigma\ \sqrt{-det(G_{ij}+\F_{ij})}  -2i\int_{M_2}\int_{0}^{1}dt\ \bar{\theta}\hL_{t}\wedge{\cal
J}L_{t}. \label{1brane}
\end{equation}

The results of applying the Killing gauge of \cite{kr} to this action
are contained in Appendix C.  This gauge was adapted to $AdS_3\times
S^3$ in \cite{rr}.  The results are not contained in the main body of
the paper as they do not simplify the action as greatly as for the
fundamental string (especially for the D5-brane actions found next).


\setcounter{equation}{0}

\section{D5-brane in $AdS_3\times S^3$}
As mentioned earlier, the action described here is for the D5-brane
lying entirely in $AdS_3\times S^3$, 
which can be interpreted as a D5-brane in $AdS_3\times S^3\times {\cal
M}^4$ with constraints imposed on the fields such that the solutions
of the action must lie in $AdS_3\times S^3$, with the ${\cal M}^4$
compactified away.  Such constraints on a D5-brane solution in
$AdS_3\times S^3\times {\cal M}^4$ describe branes in a BPS
configuration \cite{at}.  This action does not have solutions that
describe the branes whose gravity is warping spacetime to an $AdS_3\times
S^3\times {\cal M}^4$ background.  Such a D5-brane, with 4 directions
compactified in ${\cal M}^4$, would have an action like the one of
the D1-brane in $AdS_3\times S^3$ just found (\ref{1brane}).

This case of a D5-brane lying entirely in $AdS_3\times S^3$, with
${\cal M}^4$ being compactified away, has similarities to the case 
studied in \cite{abkz} where a D9-brane filling 10-dimensional space is
investigated. 

The D5-brane has a BI-action of

\begin{equation}
S_{BI}=-\int_{M_{6}} d^6\sigma\sqrt{-det(G_{ij}+\F_{ij})}
\hspace{2cm} i,j\;\;\in\;\{0,\ldots,5\}.
\end{equation}
This time the operator $\Gamma$ that defines the $\kappa$-invariance is

\begin{equation}
\Gamma=\frac{\varepsilon^{i_1 \ldots i_6}}{{\cal
L}_{BI}}\left(\frac{\Gamma_{i_1
\ldots i_6}}{6!}{\cal J} +\frac{\Gamma_{i_1
\ldots i_4}\F_{i_5 i_6}}{2.4!}{\cal E}+\frac{\Gamma_{i_1 i_2}\F_{i_3
i_4}\F_{i_5 i_6}}{2^4}{\cal J}+\frac{\F_{i_1 i_2}\F_{i_3
i_4}\F_{i_5 i_6}}{3!.2^3}{\cal E}\right). \label{gamma}
\end{equation}
Using this the same procedure as for the D1-brane is followed.

The variation $\delta_{\kappa}S_{BI}$ turns out to be given by

\begin{equation}
\delta_{\kappa}S_{BI}= 
2i\int_{M_6}\left(\frac{(\bar{ \kappa}(\hL)^5\wedge{\cal J}L)}{5!} +
\frac{(\bar{\kappa}(\hL)^3\wedge{\cal
E}L)\wedge \F}{3!}+ \frac{(\bar{\kappa}\hL\wedge{\cal
J}L)\wedge \F\wedge \F}{2}\right) \label{deltaBI}.
\end{equation}
The WZ term whose $\kappa$-variation cancels this is
\begin{equation}
S_{WZ}= -2i\int_{M_6}\int_{0}^{1}dt\ \left(\frac{(\bar{ \theta}(\hL)^5
\wedge{\cal J}L)}{5!} + \frac{(\bar{\theta}(\hL)^3\wedge{\cal
E}L)\wedge \F}{3!}+ \frac{(\bar{\theta}\hL\wedge{\cal
J}L)\wedge \F\wedge \F}{2}\right).
\end{equation}

The $\kappa$-variation of this is
\begin{equation}
\delta_{\kappa}S_{WZ}= 
-2i\int_{M_6}\left(\frac{(\bar{ \kappa}(\hL)^5\wedge{\cal J}L)}{5!} +
\frac{(\bar{\kappa}(\hL)^3\wedge{\cal
E}L)\wedge \F}{3!}+ \frac{(\bar{\kappa}\hL\wedge{\cal
J}L)\wedge \F\wedge \F}{2}\right) 
+ Y,
\end{equation}
where
$$
Y=\int_{M_6}i_\kappa\int_{0}^{1}dt\ \left(-\frac{8}{5!} (\bar{\theta}\sigma_{-}
((L_{t}^a\Gamma_a)^5\wedge L_{t}^{b'}\Gamma_{b'} - L_{t}^a\Gamma_a\wedge
(L_{t}^{b'}\Gamma_{b'})^5) \wedge {\cal K}L_{t})\right)
$$
\begin{equation}\label{Y}
+\int_{M_6}i_\kappa\int_{0}^{1}dt\ \left(
\frac{1}{3}(\bar{\theta}\sigma_{-} ((L_{t}^a\Gamma_a)^4 -
(L_{t}^{a'}\Gamma_{a'})^4) \wedge L_{t}) \wedge \F_{t} \right), \nonumber
\end{equation}
and where $i_\kappa$ is from $\delta_\kappa=\{d,i_\kappa\}$ and $i_\kappa
L=\kappa$, $i_\kappa L^{\hat a}=0$.

Due to the indices $a\ \in\ \{0,1,2\}$ and $a'\ \in\ \{3,4,5\}$ having
only 3 possibilities each, $Y=0$.  

The total effective action of a D5-brane in $AdS_3\times S^3$ is thus
given by
$$
S^{D5}=-\int_{M_{6}} d^6\sigma\sqrt{-det(G_{ij}+\F_{ij})}
$$
\begin{equation}\label{S^{D5}}
-2i\int_{M_6}\int_{0}^{1}dt\ \left(\frac{(\bar{ \theta}(\hL_{t})^5 
\wedge{\cal J}L_{t})}{5!} + \frac{(\bar{\theta}(\hL_{t})^3\wedge{\cal
E}L_{t})\wedge \F_{t}}{3!}+ \frac{(\bar{\theta}\hL_{t}\wedge{\cal
J}L_{t})\wedge \F_{t}\wedge \F_{t}}{2}\right).
\end{equation}
When calculating this the Fierz identity eqn (\ref{Fierz1}) was
required as well as identities \cite{cgnw,cgnsw,s}
\begin{eqnarray}
(\Gamma^{\hat{a}\hat{b}\hat{c}}{\cal E})_{(\hat{\alpha}
\hat{\beta}}(\Gamma_{\hat{c}})_{\hat{\gamma}\hat{\delta})}-
2(\Gamma^{[\hat{a}}{\cal
E})_{(\hat{\alpha}\hat{\beta}}(\Gamma^{\hat{b}]}{\cal
K})_{\hat{\gamma} \hat{\delta})}=0,\\
(\Gamma^{\hat{a}\hat{b}\hat{c}\hat{d}\hat{e}}{\cal J})_{(\hat{\alpha}
\hat{\beta}}(\Gamma_{\hat{e}})_{\hat{\gamma}\hat{\delta})}-
4(\Gamma^{[\hat{a}\hat{b}\hat{c}}{\cal
E})_{(\hat{\alpha}\hat{\beta}}(\Gamma^{\hat{d}]}{\cal
K})_{\hat{\gamma} \hat{\delta})}=0.
\end{eqnarray}

The Killing guage fixed D5-brane action in $AdS_3\times S^3$ is given in
Appendix C.

\setcounter{equation}{0}

\section{D5-brane in $AdS_5\times S^5$}

The mathematics in this case is almost identical to the previous case
(this is due to the intentional similarities in the notation), however
in this case $Y$ is not zero, but we can use

\begin{equation}
\epsilon^{a_1\ldots a_5}=-i\sigma_1 \Gamma^{a_1\ldots a_5},
\hspace{2cm} 
\epsilon^{a'_1\ldots a'_5}=\sigma_2 \Gamma^{a'_1\ldots a'_5},
\end{equation}
to show that

\begin{eqnarray}
S^{D5}&=&-\int_{M_{6}} d^6\sigma\sqrt{-det(G_{ij}+\F_{ij})} 
\nonumber \\ 
& &
-i\int_{M_7}\ \left( \frac{(\bar{L}\wedge (\hL)^5
\wedge {\cal J} L)}{5!} + \frac{(\bar{L} \wedge (\hL)^3 \wedge{\cal
E}L)\wedge \F}{3!}+ \frac{(\bar{L}\wedge\hL\wedge{\cal
J}L)\wedge \F\wedge \F}{2}\right)
\nonumber \\
& & +\int_{M_7}\left(\frac{\epsilon_{a_1\ldots
a_5}}{30}L^{a_1}\wedge\ldots \wedge L^{a_5}\wedge\F 
+\frac{\epsilon_{a'_1\ldots
a'_5}}{30}L^{a'_1}\wedge\ldots \wedge L^{a'_5}\wedge\F \right), \label{yippee}
\end{eqnarray}
is the full $\kappa$-invariant, supersymmetric action.

Eqn (\ref{yippee}) can be rewritten as 
$$
S^{D5}=S_{BI}-2i\int_{M_6}\int_{0}^{1}dt\ \left(\frac{(\bar{ \theta}(\hL_{t})^5
\wedge{\cal J}L_t)}{5!} + \frac{(\bar{\theta}(\hL_t)^3\wedge{\cal
E}L_t)\wedge \F_t}{3!}+ \frac{(\bar{\theta}\hL_t\wedge{\cal
J}L_t)\wedge \F_t\wedge \F_t}{2}\right)
$$
\begin{equation}\label{S^{D5}}
+\int_{M_7} {\cal L}_{WZ}^{BOSE} ,
\end{equation}
where
\begin{equation}
{\cal L}_{WZ}^{BOSE}={\cal L}_{WZ}|_{\theta=0}=
\frac{\epsilon_{a_1\ldots
a_5}}{30}e^{a_1}\wedge\ldots \wedge e^{a_5}\wedge dA 
+\frac{\epsilon_{a'_1\ldots
a'_5}}{30}e^{a'_1}\wedge\ldots \wedge e^{a'_5}\wedge dA.
\label{bose}
\end{equation}

The Killing guage fixed D5-brane action in $AdS_5\times S^5$ is
contained in
Appendix C.

\setcounter{equation}{0}

\section{D5-brane in $AdS_3\times S^3\times T^4$}

As mentioned in the introduction, an attempt was made to find the
action of a
general D5-brane free to move in all the dimensions of
$AdS_3\times S^3\times T^4$.  The same method as used to find the
other actions in this paper is used, however as will be shown, this
method does not succeed in producing a fully $\kappa$-invariant action
in this case.

The case of extending the D1-brane
action to $AdS_3\times S^3\times T^4$ is trivial, and in the notation
of Appendix A, appears unchanged to (\ref{1brane}).  

As mentioned in Appendix A, the coset representation for $T^4$ that is
used here describes only translations in
$T^4$ (ie $U(1)^4$).

The D5-brane BI-action is

\begin{eqnarray}
& &S_{BI}=-\int_{M_{6}} d^6\sigma\sqrt{-det(G_{ij}+\F_{ij})},\hspace{2cm}
i,j\ \in \ \{0,\ldots,5\},
\nonumber \\
& &G_{ij}=L_{i}^a L_{j a} + L_{i}^{a'} L_{j a'}+L_{i}^{a''} L_{j a''}, \\
& &\F=dA+2i\int_{0}^1 dt\ \bar{\theta}(L_{t}^a \Gamma_{a} + L_{t}^{a'}
\Gamma_{a'}+L_{t}^{''}\Gamma_{a''}){\cal K}\wedge L_t,   
\nonumber
\end{eqnarray}
where $\Gamma$ in given by (\ref{gamma}).

The $\kappa$-variation of the BI-action appears the same as (\ref{deltaBI}) in
the $AdS_3\times S^3\times T^4$ notation of Appendix A.

The WZ-action required to cancel this should be of the form

\begin{eqnarray}
S^{D5}_{WZ} &=& -2i\int_{M_6}\int_{0}^{1}dt\ \left(\frac{(\bar
{\theta}(\hL_t)^5 
\wedge{\cal J}L_t)}{5!} + \frac{(\bar{\theta}(\hL_t)^3\wedge{\cal
E}L_t)\wedge \F_t}{3!} 
 +\frac{(\bar{\theta}\hL_t\wedge{\cal
J}L_t)\wedge \F_t\wedge \F_t}{2}\right)
\nonumber \\
& & +\int_{M_7} E_{7}^{BOSE} 
\\
&=& -i \int_{M_7}\ \left(\frac{(\bar{L}\wedge (\hL)^5
\wedge {\cal J}L)}{5!} + \frac{ ( \bar{L} \wedge (\hL)^3 \wedge {\cal
E} L) \wedge \F}{3!}+ \frac{(\bar{L} \wedge \hL \wedge {\cal
J} L)\wedge \F\wedge \F}{2} \right) 
\nonumber \\ 
& &+\int_{M_7} E_{7},
\end{eqnarray}
where $E_7$ should have the following properties to make the action
$\kappa$-invariant 
\begin{itemize}
\item $E_7$ should be supersymmetric.
\item $i_\kappa E_7=0$.
\item 
$$
\delta_\kappa E_7=
\frac{8}{5!}(\bar{\kappa}\sigma_{-}
((L^a\Gamma_a)^5\wedge L^{\tilde{b}}\Gamma_{\tilde{b}} - L^a\Gamma_a\wedge
(L^{\tilde{b}}\Gamma_{\tilde{b}})^5
) \wedge {\cal K}L)
$$
\begin{equation}
-\frac{1}{3}(\bar{\kappa}\sigma_{-} ((L^a\Gamma_a)^4 -
(L^{\tilde{a}}\Gamma_{\tilde{a}})^4) \wedge L) \wedge \F  
\end{equation}
\end{itemize}
where $\tilde{a} \in \{ a', a'' \}$.

No supersymmetric 7 form can be constructed out of the Cartan forms
and $F$ that satisfies these conditions.

\section{Summary}

In this paper the fully $\kappa$-invariant and supersymmetric actions
were found for the D1 and D5-branes in
$AdS_3\times S^3$ and the D5 action in $AdS_5\times S^5$.  The 
solutions of the D5 action in $AdS_3\times S^3$ are a BPS system as
they are S-dual to the
two intersecting NS-five branes of \cite{at}, which are BPS.  This
D5-brane has no dimensions compactified on ${\cal M}^4$.  A D5-brane that
is parallel to the branes that are the source of the background field
would have 
four dimensions compactified on ${\cal M}^4$ and the action would be
the same as the D1-brane action.

The D5-brane action in $AdS_5\times S^5$ that was found is of interest
as it can be adjusted to describe the D5, D3, F1 configuration of
\cite{cgmp,ew2,grst,yi,4p} which exhibits a baryon vertex, by placing some
restrictions on the degrees of freedom.

A fully $\kappa$-symmetric D5 action in $AdS_3\times S^3\times T^4$
failed to be found in this case, however I do not
believe that this is a reason to believe that it does not exist.  The
WZ-action may take a form very different to more standard D-brane
actions.  It should be said though that most of the interesting
dynamics occurs in the $AdS_3\times S^3$, not in $T^4$, so the action
found in (\ref{1brane}) is sufficient for most purposes.

In Appendix C the Killing gauge was applied to the actions, however
even though the D1-brane and D5-brane actions in $AdS_3\times S^3$ are
reduced to being eigth order in fermions, they are still somewhat
complicated.  The D5-brane in $AdS_5\times S^5$ remains very
complicated even in this gauge.  It remains to be seen if some method,
such as the T-duality transformation of \cite{kt} can be used to
simplify the actions. 

\vspace{12mm}
\begin{flushleft}
{\Large {\bf Acknowledgements}}
\end{flushleft}

I would like to thank my supervisor Peter Bouwknegt for providing
useful guidance and R.R. Metsaev and A.A. Tseytlin
for several invaluable correspondences.  My research has been
supported by an Australian Postgraduate Award scholarship.

\oddsidemargin 0mm


\setcounter{equation}{0}

\appendix
\section{Notation and Algebra for Various
$AdS\times S$ Spaces}

The algebra follows along the lines of \cite{p,rr,pr,mt}.  

\subsection{$AdS_3\times S^3\times T^4$}

The coset
representation
for $T^4$ that is used here describes only translations.
That is, $J_{a'' b''}$ are absent from the algebra.  It
is not certain if this situation was described in an ideal fashion.

The notation and algebra used 
(with the radii of the spaces $AdS_3$, $S^3$ and $T^4$ set to 1) is as
follows  

\begin{itemize}

\item $a,b,c\; \in\, \{0,1,2\}$ are indices of $SO(2,1)$ ($AdS_3$).
\item $a',b',c'\; \in\, \{3,4,5\}$ are indices of $SO(3)$ ($S^3$).
\item $a'',b'',c''\; \in\, \{6,7,8,9\}$ are indices of $U(1)^4$
($T^{4}$). 
\item $\hat{a},\hat{b},\hat{c}\; \in\, \{0,\ldots,9\}$ are a combination of
$(a,b,c)$, $(a',b',c')$ and $(a'',b'',c'')$.
\item $\alpha,\beta,\gamma\; \in\, \{1,2\}$ are $SO(2,1)$ spinor indices.
\item $\alpha',\beta',\gamma'\;\in\, \{1,2\}$ are $SO(3)$ spinor
indices.
\item $\alpha'',\beta'',\gamma''\;\in\,\{1,\ldots,4\}$ are $SO(4)$
spinor indices. 
\item $\hat{\alpha},\hat{\beta},\hat{\gamma}\; \in\, \{1,\dots,32\}$
are $D=10$ Majorana-Weyl spinor indices.
\item $I,J,K \;\in\, \{1,2\}$ are ${\cal N} = 2$ supersymmetry labels.
\item $(P_{a},J_{a b})$, $(P_{a'},J_{a' b'})$ and
$(P_{a''},J_{a'' b''})$ are the $SO(2,1)$,
$SO(3)$ and $SO(4)$ translations and rotations
respectively.  They are defined to be antihermitian.
\item $32\times 32$ $\Gamma$ matrices of $AdS_{3}\times S^{3}\times T^{4}$
are \newline
\begin{eqnarray}
& &\Gamma^{a}=\gamma^{a}\otimes 1_{2}\otimes 1_{4}\otimes\sigma^{1},\ \ \   
\gamma^{a}:\ \ \gamma^{0}=i\sigma^{3},\ \  \gamma^{1}=\sigma^{1},
\ \ \gamma^{2}=\sigma^{2},\\
& &\Gamma^{a'}=1_{2} \otimes\gamma^{a'}\otimes{\bar
\gamma}_{5}\otimes\sigma^{2},\ \ \  
\gamma^{a'}:\ \ \gamma^{3'}=\sigma^{1},\ \  \gamma^{4'}=\sigma^{2},\ 
\ \gamma^{5'}=\sigma^{3}, \ \ \ {\bar
\gamma}_{5}=\gamma_{6}\gamma_{7}\gamma_{8}\gamma_{9}, \nonumber \\
& &\Gamma^{a''}=1_{2} \otimes
1_{2}\otimes\gamma^{a''}\otimes\sigma^{2}.
\nonumber 
\end{eqnarray}
\item 
$ \pi_+ = 1_2\otimes 1_2\otimes\ 1_4 \otimes \frac{1}{2}(1_2 +
\sigma_3),\ \ 
\sigma_+=1_2\otimes 1_2\otimes\ 1_4 \otimes
\frac{1}{2}(\sigma_1+i\sigma_2)$\\  
$\sigma_-=1_2\otimes 1_2\otimes\ 1_4 \otimes
\frac{1}{2}(\sigma_1-i\sigma_2)$
\item ${\bf {\hat C}}=C\otimes C'\otimes\ C''\otimes i\sigma^{2}$ is the charge
conjugation matrix, where $C$, $C'$ and $C''$ are the charge conjugation
matrices of $SO(2,1)$, $SO(3)$ and $SO(4)$.
\item $\{\Gamma^{\hat{a}},\Gamma^{\hat{b}}\}=2\eta^{\hat{a} \hat{b}}$
where $\eta^{\hat{a} \hat{b}}=(-,+,\ldots,+)$. 
\item Supersymmetry generators $Q^{I
\hat{\alpha}}=\left(\begin{array}{c}0\\-Q^{I \alpha \alpha'
\alpha''}\end{array}\right)$.  In this notation $\theta_{I
\hat{\alpha}}$ and $L_{I\hat{\alpha}}$ have the opposite chirality:
$L^{I \hat{\alpha}}=\left(\begin{array}{c}L^{I \alpha \alpha'
\alpha''}\\0 \end{array}\right)$.  
\item $Q^{1\hat{\alpha}}$ and $Q^{2\hat{\alpha}}$ can be combined as
$Q=\left(\begin{array}{c}Q^{1}\\Q^{2}\end{array}\right)$.
\item 
\begin{equation}
{\cal E}=\left(\begin{array}{cc} 0 & 1\\-1 &
0\end{array}\right)\;\;\;
{\cal J}=\left(\begin{array}{cc} 0 & 1\\1 & 0\end{array}\right)\;\;\;
{\cal K}=\left(\begin{array}{cc} 1 & 0\\0 & -1\end{array}\right) {\rm
\ act\ on\ } Q=\left(\begin{array}{c}Q^{1}\\Q^{2}\end{array}\right).
\label{Q}
\end{equation}

\end{itemize}

The algebra

\begin{eqnarray}
& &[P_a,P_b]=J_{ab},\hspace{2cm} [P_{a'},P_{b'}]=-J_{a'b'},\hspace{2cm}
[P_{a''},P_{b''}]=0,
\nonumber \\ {} 
& &[P_{a},J_{bc}]=\eta_{ab}P_{c}-\eta_{ac}P_{b},\hspace{2cm} 
[P_{a'},J_{b'c'}]=\eta_{a'b'}P_{c'}-\eta_{a'c'}P_{b'},
\nonumber \\ {}
& &[J_{ab},J_{cd}]=\eta_{bc}J_{ad}+\eta_{ad}J_{bc}-\eta_{ac}J_{bd}
-\eta_{bd}J_{ac},
\\ {}
& &[J_{a'b'},J_{c'd'}]=\eta_{b'c'}J_{a'd'}+\eta_{a'd'}
J_{b'c'}-\eta_{a'c'} J_{b'd'} -\eta_{b'd'}J_{a'c'},
\nonumber \\ {}
& &[Q,P_{\hat a}]=\frac{i}{2}Q{\cal E}\sigma_+\Gamma_{\hat a}, \hspace{2cm} 
[Q,J_{{\hat a}{\hat b}}]=-\frac{1}{2}Q\Gamma_{{\hat a}{\hat b}}\ \ {\rm
if}\ \ {\hat a},\ {\hat b}\ \leq 5,
\nonumber \\ {}
& &\{Q_{\hat \alpha},Q_{\hat \beta}\}= -2i({\hat C}\Gamma^{\hat
a}\pi_+)_{{\hat \alpha}{\hat \beta}}P_{\hat a}+ {\cal E}[({\hat
C}\Gamma^{ab}\sigma_-)_{{\hat \alpha}{\hat \beta}}J_{ab}-({\hat
C}\Gamma^{a'b'}\sigma_-)_{{\hat \alpha}{\hat \beta}}J_{a'b'}].
\nonumber
\end{eqnarray}

\subsection{$AdS_3\times S^3$}

The notation and algebra for just $AdS_3\times S^3$ space (ie,
assuming the $T^4$ is compactified to zero volume limit and ignoring
it) is almost identical to the case of the more general $AdS_3\times
S^3\times T^4$, except any terms with double primed indices are absent,
thus any difficulties in treating 
The
notable changes besides this are

\begin{itemize} 
\item $\hat{\alpha},\hat{\beta},\hat{\gamma}\; \in\, \{1,\dots,8\}$
are $D=6$ 
complex chiral
spinor indices.
\item $8\times 8$ $\Gamma$ matrices of $AdS_{3}\times S^{3}$ are \newline
\begin{eqnarray}
\Gamma^{a}&=&\gamma^{a}\otimes 1\otimes\sigma^{1},\ \ \   
\gamma^{a}:\ \ \gamma^{0}=i\sigma^{3},\ \  \gamma^{1}=\sigma^{1},
\ \ \gamma^{2}=\sigma^{2},\\
\Gamma^{a'}&=&1 \otimes\gamma^{a'}\otimes\sigma^{2},\ \ \  
\gamma^{a'}:\ \ \gamma^{3'}=\sigma^{1},\ \  \gamma^{4'}=\sigma^{2},\ 
\ \gamma^{5'}=\sigma^{3}.
\nonumber
\end{eqnarray}
\end{itemize}

\subsection{$AdS_5\times S^5$}

The third set of notation and algebra we use is for $AdS_5\times
S^5$.  This is exactly the same notation as used in \cite{mt}.

\begin{itemize}

\item $a,b,c\; \in\, \{0,\ldots,4\}$ are indices of $SO(4,1)$.
\item $a',b',c'\; \in\, \{5,\ldots,9\}$ are indices of $SO(5)$.
\item $\hat{a},\hat{b},\hat{c}\; \in\, \{0,\ldots,9\}$ are a combination of
$(a,b,c)$ and $(a',b',c')$.
\item $\alpha,\beta,\gamma\; \in\, \{1,\ldots,4\}$ are $SO(4,1)$
spinor indices. 
\item $\alpha',\beta',\gamma'\;\in\, \{1,\ldots,4\}$ are $SO(5)$ spinor
indices.
\item $\hat{\alpha},\hat{\beta},\hat{\gamma}\; \in\, \{1,\dots,32\}$
are $D=10$ Majorana-Weyl spinor indices.
\item $I,J,K \;\in\, \{1,2\},\; {\cal N} = 2$ supersymmetry labels.
\item $(P_{a},J_{a b})$, $(P_{a'},J_{a' b'})$ and are the $SO(4,1)$ and
$SO(5)$ translations and rotations respectively.  They are
defined to be antihermitian.
\item $32\times 32$ $\Gamma$ matrices of $AdS_{5}\times S^{5}$
are \newline
\begin{eqnarray}
\Gamma^{a}&=&\gamma^{a}\otimes 1_{4}\otimes\sigma^{1},\ \ \   
\gamma^{a}:\ \ \ \gamma^0=i\sigma^3\otimes 1_2,\ \
\gamma^{1,\ldots,3}=\sigma^2\otimes \sigma^{1,\ldots,3},\ \
\gamma^4=\sigma^1\otimes 1_2, \nonumber\\ 
\Gamma^{a'}&=&1_{4} \otimes\gamma^{a'}\otimes\sigma^{2},\ \ \  
\gamma^{a'}:\ \ \ \gamma^{5,\ldots,7}=\sigma^2\otimes
\sigma^{1,\ldots,3},\ \ \gamma^8=\sigma^3\otimes 1_2,\ \ \gamma^9=
\sigma^1 \otimes 1_2.
\nonumber
\\
\end{eqnarray}
\item ${\bf {\hat C}}=C\otimes C'\otimes\ i\sigma^{2}$ is the charge
conjugation matrix, where $C$ and $C'$ are the charge conjugation
matrices of $SO(4,1)$ and $SO(5)$.
\item $\{\Gamma^{\hat{a}},\Gamma^{\hat{b}}\}=2\eta^{\hat{a} \hat{b}}$
where $\eta^{\hat{a} \hat{b}}=(-,+,\ldots,+)$. 
\item Supersymmetry generators $Q^{I
\hat{\alpha}}=\left(\begin{array}{c}0\\-Q^{I \alpha
\alpha'}\end{array}\right)$.  In this notation $\theta_{I
\hat{\alpha}}$ and $L_{I\hat{\alpha}}$ have the opposite chirality: 
$L^{I \hat{\alpha}}=\left(\begin{array}{c}L^{I \alpha \alpha'
}\\0 \end{array}\right)$.
\item $Q^{1\hat{\alpha}}$ and $Q^{2\hat{\alpha}}$ can be combined as
$Q=\left(\begin{array}{c}Q^{1}\\Q^{2}\end{array}\right)$.
\end{itemize}


\setcounter{equation}{0}

\section{Maurer-Cartan Equations and other Cartan form relations}

The Maurer-Cartan Equations, the variations for the Cartan forms and
their $\partial_t L$ equations (which will be explained below) for
$AdS_3\times S^3\times T^4$, 
$AdS_3\times S^3$ and $AdS_5\times S^5$ can all be written out in the
same form, as long as it is understood that the indices run over
different ranges for each space, and that $L^{a'' b''}=0$ for
$AdS_3\times S^3 \times T^4$.  The equations of course appear exactly
the same as the corresponding equations in \cite{mt}.
\begin{eqnarray}
& &dL^{\hat{a}}=-L^{\hat{a}\hat{b}}\wedge L_{\hat{b}}-i{\bar
L}\Gamma^{\hat{a}}\wedge L \label{dL1}\\
& &dL=\frac{i}{2}\sigma_{+}\hL\wedge{\cal E} L-\frac{1}{4}
L^{\hat{a}\hat{b}} \Gamma_{\hat{a}\hat{b}}\wedge L \label{dL2}\\
& &dL^{a b}=-L^{a}\wedge L^{b} -L^{ac}\wedge L_{c}^{\ \ b}+
\bar{L}\Gamma^{ab}\sigma_{-}\wedge{\cal E} L\\
& &dL^{a' b'}=\L^{a'}\wedge L^{b'} -L^{a'c'}\wedge L_{c'}^{\ \ b'}-
\bar{L}\Gamma^{a'b'}\sigma_{-}\wedge{\cal E} L
\end{eqnarray}

The variations of the Cartan forms are

\begin{eqnarray}
& &\delta L^{\hat a}=d\delta x^{\hat a} + L^{\hat{a}\hat{b}}\delta x_{\hat b} -
\delta x^{\hat{a}\hat{b}}L_{\hat b}+2i\bar{L}\Gamma^{\hat a}\delta
\theta \label{varL1},\\
& &\delta L= d\delta\theta
- \frac{i}{2}\sigma_{+}\hL{\cal E}\delta\theta 
 + \frac{1}{4} L^{\hat{a}\hat{b}} \Gamma_{\hat{a}\hat{b}} \delta \theta
 +\frac{i}{2} \sigma_{+}\delta x^{\hat{a}}\Gamma_{\hat a}{\cal E} L-
 \frac{1}{4} \delta x^{\hat{a}\hat{b}}\Gamma_{\hat{a}\hat{b}}L, \label{varL2}\\
& &\delta x^{\hat a}\equiv \delta X^{M}L^{\hat a}_{M},\ \ \ \delta
 x^{\hat{a}\hat{b}} \equiv \delta X^{M} L^{\hat{a}\hat{b}}_{M},\ \ \
 \delta \theta\equiv\delta X^{M} L^{\hat{a}\hat{b}}_{M}. \label{varL3}
\end{eqnarray}

Rescale $\theta \rightarrow t\theta$ to get the $L_{t}$ Cartan forms.

\begin{equation}
L^{\hat a}_t(x,\theta) \equiv L^{\hat a}(x,t\theta),\ \ \ 
L^{\hat{a}\hat{b}}_t(x,\theta) \equiv L^{\hat{a}\hat{b}}(x,t\theta), \ \ \ 
L_t(x,\theta) \equiv L(x,t\theta). 
\end{equation}

The initial conditions of these are
\begin{equation}
L_{t=0}^{\hat a}=e^{\hat a},\hspace{2cm} L_{t=0}^{{\hat a}{\hat
b}}=\omega^{{\hat a}{\hat b}},\hspace{2cm} L_{t=0}=0. \label{initial}
\end{equation}

It can be shown that
\begin{equation}
\partial_t \F_t=2i \bar{\theta}\hL_t\wedge{\cal K} L_{t}.
\end{equation}

The defining equations for the Cartan forms can be shown to be \cite{mtfirst}

\begin{eqnarray}
& &\partial_{t} L^{\hat a}_{t}=-2i\bar{\theta}\Gamma^{\hat
a}L_{t},\label{pL1}\\
& &\partial_{t} L_{t}=d\theta-\frac{i}{2}\sigma_{+}\hL{\cal
E}\theta + \frac{1}{4}\Gamma_{\hat{a}\hat{b}}L^{\hat{a}\hat{b}}_{t},
\label{pL2} \\
& &\partial_{t} L^{ab}_{t}=2\bar{\theta}{\cal
E}\Gamma^{ab}\sigma_{-}L_{t},\hspace{2cm} \partial_{t}
L^{a'b'}_{t}=-2\bar{\theta}{\cal E}\Gamma^{a'b'}\sigma_{-}L_{t}. 
\end{eqnarray}

The solutions \cite{krr} to these equations can be found in equations
B.14 - B.18 of \cite{mt} and using the Killing gauge they are
symplified in \cite{kr,ppdsmt}.  

\setcounter{equation}{0}

\section{Killing Gauge Fixed Actions}

In this appendix the Killing gauge of \cite{kr} is applied to the
D1-brane in $AdS_3\times S^3$, and the D5-brane actions in both
$AdS_3\times S^3$ and $AdS_5\times S^5$.

\subsection{D1-brane in $AdS_3\times S^3$}

The Killing gauge \cite{kr,rr} is defined by

\begin{eqnarray}
{\cal P}_{\pm}^{IJ}=\frac{1}{2}(\delta^{IJ} \pm \Gamma_{*}
\epsilon^{IJ}),\\
{\cal P}_- \theta=0,
\end{eqnarray}
where $\Gamma_{*}=i\Gamma_{01}$ in $AdS_3\times S^3$.

The basis used to represent the spinors in the simplified
action is

\begin{eqnarray}
\theta_{\pm}={\cal P}_{\pm}\theta,\\
\theta_{\pm}^{1}=\frac{1}{2}(\theta^1 \pm \Gamma_* \theta^2),\\
\theta_{\pm}^{2}=\frac{1}{2}(\theta^2 \mp \Gamma_* \theta^1) =
\mp\Gamma_* \theta_{\pm}^1.
\end{eqnarray}

Letting $p\in {0,1}$ be the coordinates along the background D1-branes
and $q \in \{2,\ldots,5\}$ be the transverse coordinates, $\theta$ and
the Cartan forms can be shown to simplify to

\begin{eqnarray}
\theta_{+}^I=\sqrt{|y|}\eta_{+}^I,\\
L_{t\ +}^I=t\sqrt{|y|}d\eta_{+}^I,\\
L_{-}^I=0,\\
L^{p}_t=|y|(dx^p-it^2{\bar \eta}_{+}^I \Gamma^p d\eta_{+}^I),\\
L_{t}^q=\frac{1}{|y|}dy^q.
\end{eqnarray}
Using this, the BI-action is

\begin{eqnarray}
& &S_{BI}=-\int_{M_{2}} d^2\sigma\ \sqrt{-det(G_{ij}+\F_{ij})} \label{last},\\
& &G_{ij}={\my}^2 (\p_i x_p-2i (\nl\Gamma_p \p_i\nr)) (\p_j x^p-2i
(\nl\Gamma^p \p_j\nr)) + \frac{1}{\my^2}dy^q dy^q,\\
& &\epsilon^{ij} F_{ij} = \epsilon^{ij} (2\p_{i}A_{j}+ 4i(\nl
\Gamma_q \p_{i}y^q \p_{j} \nr)), 
\end{eqnarray}
while the WZ-action is found to be
\begin{equation}
S_{WZ}=-2\int_{M_2} (\nl (dy^q \Gamma_q)\Gamma_{01}\wedge d\nr).
\end{equation}


\subsection{D5-brane in $AdS_3\times S^3$}

The BI-action of the D5-brane looks identical to that of the D1-brane,
except of course that $i, j \in \{0,\ldots, 5\}$ instead of
$\{0,1\}$.

The WZ-action is quite complicated in this gauge, however some terms
disappear as the Killing gauge leaves only 8 fermionic degrees of
freedom, so no term can be higher than eighth order in $\nr$.


\begin{eqnarray}
& &S_{WZ} = -2i\int_{M_6} \left[-\frac{2}{5!}\nl\left(\frac{1}{2\my^4}
(dy^q \Gamma_q)^5 
+10(dy^q\Gamma_q)^3 \wedge 
\left( \frac{1}{2}(dx^p\Gamma_p)^2 \right.\right.\right. \nonumber\\
& &\left.\left.\left.-i(dx^p\Gamma_p) \wedge \nn 
 -\frac{2}{3} \nn^2 \right) 
+ 5\my^5
(dy^q\Gamma_q) \wedge \left( \frac{1}{2}(dx^p\Gamma_p)^4
\right.\right.\right. \nonumber \\
& &\left.\left.\left. -2i\dx^3
\wedge \nn -4\dx^2\wedge\nn^2 +4i\dx\wedge
\right.\right.\right. \nonumber\\
& &\left.\left.\left. \ \nn^3 
\right)\right) \wedge \Gamma_* d\nr \right. \nonumber \\
& &\left. +
\frac{1}{3!}\nl\left(-2\my^4\left(\frac{1}{2}\dx^3-\frac{3i}{2}\dx^2 
\wedge \nn -2 \dx\wedge \nn^2 \right.\right.\right.\nonumber \\
& &\left.\left.\left.  +i\nn^3\right) -3\dx\wedge\dy^2+
3i\dy^2\wedge \nn \right)\wedge \Gamma_* d\nr \wedge dA\right. \nonumber\\
& &\left. + \frac{i}{3}\nl 
\left(-2\my^4\left(\frac{1}{4}\dx^3-i\dx^2\wedge \nn -\frac{3}{2}\dx
\wedge\nn^2 \right)\right.\right. \nonumber\\
& &\left.\left. 
 -\frac{3}{2}\dx\wedge \dy^2
+2i\nn\wedge\dy^2 \right) \wedge \Gamma_* d\nl \wedge
\right. \nonumber\\
& &\left. \ (\nl\dy\wedge d\nr) 
-(\nl\dy\wedge\Gamma_* d\nr)\wedge \left(\frac{1}{2}dA\wedge
dA + idA\wedge (\nl\dy\wedge d\nr) \right.\right.\nonumber\\
& &\left.\left.
-\frac{2}{3}(\nl\dy\wedge d\nr)^2\right) 
\right]. 
\end{eqnarray}

\subsection{Killing Gauge Fixed D5-brane Action in $AdS_5\times S^5$}

Again the BI-action is written as eqn (\ref{last}) with $i,j\in\{0,
\ldots, 5\}$, and the WZ-action is found to be very similar to the $AdS_3\times
S^3$ case, except that now $\Gamma_*=\Gamma_{0123}$ and the WZ-action
contains some 
extra terms due to the greater number of fermionic degrees of freedom,
and also due to eqn (\ref{bose}).

\begin{eqnarray}
& &S_{WZ} = -2i\int_{M_6} \left[-\frac{2}{5!}\nl\left(\frac{1}{2\my^4}
(dy^q \Gamma_q)^5 
+10(dy^q\Gamma_q)^3 \wedge 
\left( \frac{1}{2}(dx^p\Gamma_p)^2 \right.\right.\right. \nonumber\\
& &\left.\left.\left.-i(dx^p\Gamma_p) \wedge \nn 
 -\frac{2}{3} \nn^2 \right) 
+ 5\my^5
(dy^q\Gamma_q) \wedge \left( \frac{1}{2}(dx^p\Gamma_p)^4
\right.\right.\right. \nonumber \\
& &\left.\left.\left. -2i\dx^3
\wedge \nn -4\dx^2\wedge\nn^2 +4i\dx\wedge
\right.\right.\right. \nonumber\\
& &\left.\left.\left. \ \nn^3 + \frac{8}{5}\nn^4
\right)\right) \wedge \Gamma_* d\nr \right. \nonumber \\
& &\left. +
\frac{1}{3!}\nl\left(-2\my^4\left(\frac{1}{2}\dx^3-\frac{3i}{2}\dx^2 
\wedge \nn -2 \dx\wedge \nn^2 \right.\right.\right.\nonumber \\
& &\left.\left.\left.  +i\nn^3\right) -3\dx\wedge\dy^2+
3i\dy^2\wedge \nn \right)\wedge \Gamma_* d\nr \wedge dA\right. \nonumber\\
& &\left. + \frac{i}{3}\nl 
\left(-2\my^4\left(\frac{1}{4}\dx^3-i\dx^2\wedge \nn -\frac{3}{2}\dx
\wedge\nn^2 \right.\right.\right. \nonumber\\
& &\left.\left.\left. +\frac{4i}{5}\nn^3 \right) -\frac{3}{2}\dx\wedge \dy^2
+2i\nn\wedge\dy^2 \right) \wedge \Gamma_* d\nl \wedge
\right. \nonumber\\
& &\left. \ (\nl\dy\wedge d\nr) 
-(\nl\dy\wedge\Gamma_* d\nr)\wedge \left(\frac{1}{2}dA\wedge
dA + idA\wedge (\nl\dy\wedge d\nr) \right.\right.\nonumber\\
& &\left.\left.
-\frac{2}{3}(\nl\dy\wedge d\nr)^2\right) 
\right] + \int_{M_7} {\cal L}_{WZ}^{BOSE},\\
& &{\cal L}_{WZ}^{BOSE}=4\my^3 dx^{0}\wedge\ldots\wedge dx^{3}\wedge dy^4
\wedge dA +\frac{4}{\my^5}dy^5\wedge \ldots\wedge dy^9\wedge dA.
\end{eqnarray}



\end{document}